# Towards the B-TAMBiT: A Back-Translation with an Adjudicator with Mono and Bilingual Tests.

Mahamadou Kante*, Euloge François Kouame, Macire kante

*mahamadou.kante@uvci.edu.ci

Researchers have turned to various disciplines in search for theories that can contribute in different ways towards Information privacy. The data collection instrument (questionnaire) of these theories is in English. Nevertheless, issues related to Social Network Sites are meant for various groups with different cultural background. Therefore, cross-cultural and international studies are used in majority to address issues facing these platforms. Henceforth, there is a need to translate these instruments into other languages such as French. In this paper, we produced a mixed method for English instrument translation into French using different techniques from different approaches, the B-TAMBiT.

## 1 Introduction

The conceptualisation of the phenomenon and the investigation of probable explanations concerning Information Privacy has seen exponential development, especially with the rise of Social Networking Sites (SNSs). Researchers have turned to various disciplines in search for theories that can contribute in different ways towards the problem (Wirth, 2018). These include social theories, behavioural economics, psychology, and even quantum theory (Wirth, 2018; Kokolakis, 2017).

The most used models in privacy research from the year 2000 to 2018 (Wirth, 2018) are: (1) Privacy Calculus; (2) The Social Exchange Theory; (3) The Protection Motivation Theory; (4) The Communication Privacy management Theory; (5) Elaboration likelihood Model.

However, the data collection instrument (questionnaire) of these models are in English. For other speaking languages especially French, these instruments need to be translated (Kante, Chepken, & Oboko, 2017). Issues related to SNSs are meant for various groups with different cultural background. Therefore, cross-cultural and international studies are used in majority to address issues facing these platforms. There is a need to translate research instrument into languages that would enhance the reliability of the studies. That would considerably improve the quality of the data collected for better generalization of the obtained results.

From an extensive literature review done following the framework proposed by researchers Levy and Ellis, (2006). which is built on the guidelines of (Webster & Watson, 2002), factors that influence Self-Disclosure on Social Networking Sites were identified. The Privacy calculus theory is the base of our proposed model on self-disclosure on Social Networking Sites. We needed to collect data in an online survey targeting Facebook users. The instrument that is in English was partially adapted from Krasnova, Spiekermann, Koroleva, and Hildebrand, (2010). Within the literature, different approaches exist in translating survey instrument into other languages (Maneesriwongul & Dixon, 2004).

Sinaiko and Brislin, (1973). recommended one or more of the following techniques: (1) back-translation; (2) bilingual techniques; (3) committee approach; and (4) pretest. Back translation is an iterative process that consists of translating a target language version back into the source language version in order to verify translation of the research instrument (Pan & de La Puente, 2005). The bilingual approach consists of testing both target and source language versions among bilingual subjects in order to detect any differences in their responses (Maneesriwongul & Dixon, 2004). In the committee approach, the researcher uses a team of bilingual people to translate from the source to the target language. In pretest technique, a pilot study has to be undertook after instrument translation is completed in order to ensure that future users of the target language version can comprehend all questions and procedures (Pan & de La Puente, 2005; Maneesriwongul & Dixon, 2004). Questions considered in this paper are:

(1) What are the different approaches for translating survey instrument?
(2) Which approach can we use to translate effectively the instrument into French?
(3) What can we learn from that?

We investigated those questions with the aim of producing a French equivalence of our instrument that preserve its meaning, style and that is linguistically and semantically correct. The rest of this paper is organised as follow. First, we describe the materials and different methods we used that form our proposed approach. On this basis, we present our results followed by a discussion. Finally, we conclude in section

**2 Materials and methods**

We used a **B**ack-**T**ranslation approach with an **A**djudicator and **M**onolingual and **B**ilingual **T**ests (B-TAMBiT). That choice was driven by the relatively low cost of back-translation and it has proven its reliability when used carefully (Pan & de La Puente, 2005) and availability of translators. Though Pan and de La Puente, (2005). have suggested the use of a committee approach for instrument translation, Sinaiko and Brislin, (1973). pointed out that it is weak, because it does not necessarily control for shared misconceptions. A committee participant may be reluctant to criticize another participant suggestion. Additionally, multiple translators may work together or separately. It is important to note that in back-translation, modification of constructs and words without changing the semantic of the constructs or words that have no clear equivalence in the other language is allowed. Combining different techniques, we proceeded as follow:

*2.1 Translation*

The translation staff consisted of two independent translators. The two were bilingual in English and French. Prior to the start of the process, the purpose of the translation were explained to them. Additionally, they were given a definition of key terms used in the instrument:

- Social networking sites (SNS) are online environments in which people create self-descriptive profiles and then make links with other people they know on the site (i.e., creating a network of personal connections).
- Convenience of Maintaining Relationships: The value users derive from being able to efficiently and easily stay in touch with each other on SNSs (Krasnova et al., 2010).
- New Relation Building: the value users derive from being able to build up new connections to others on SNSs (Krasnova et al., 2010).
- Enjoyment: the satisfaction users get from having pleasant and enjoyable experiences on SNSs.
- Privacy Concern: an individual's subjective views of fairness within the context of information privacy (Campbell, 1997).
- Information sensitivity: information that contain high risk (as opposed to low risk) that makes the discloser feel vulnerable in some way (Moon, 2000).
- Self-disclosure: extent of information a user provides in the process of participation on an SNS (e.g. on the profile, in the process of communication with others) (Krasnova et al., 2010).

The first translator translated from the source language to the target language. Afterwards, the translation from the target language to the source language i.e. from French to English.

*2.2 Comparison*

During the translation process, the translators documented the challenges faced during the process. The two versions were compared in search for discrepancies and solutions were suggested.

*2.3 Validation*

After the translation and the comparison, a first document was then accepted and validated by a Professor of Sociology and Anthropology, highly skilled in social science researches and monolingual in French. Afterwards, an adjudicator, researcher in ICT4D bilingual in French and English went through the two versions of the instrument and validated them in turn with minor correction to adjust the two versions.

**3 Tests**

Five voluntary participants were asked to describe how they understood the instrument. Two were bilingual and completed both the instrument in English and in French and the other three (monolingual) completed the French version. Overall, no major differences were found between the two versions of the instrument. However, a final version was made based on suggestions of the respondents.

**4 Results**

From the process described above, we could produce a survey instrument that could be filled by both an English and French speaker. A number of different techniques were used in order to produce the French equivalence of our instrument that preserve its meaning, style and that is linguistically and semantically correct. Two main issues occurred while translating:

- The choice of the right corresponding word: assuming that one English could have three to four corresponding words, it was sometimes difficult to choose. In fact, researchers Kante et al., (2017) encountered the same issue in their study. Our choices were made based on suggestions from the adjudicator and from the results of the tests we conducted.

- In order to keep the translated version free of grammatical errors, the structure of some sentences were changed. However, as a constraint of this endeavour, all sentences kept their original meanings.

We believe that an effective translated instrument was formed after a rigorous process that borrows techniques from different recognize techniques. That has resulted to the B-TAMBiT approach. Back translation used as described in (Maneesriwongul & Dixon, 2004) combine with one of strength of the committee approach namely an adjudicator has proven to be effective and at low or no cost. Additionally, the use of monolingual and bilingual tests as pretests permits to detect earlier discrepancies that might have been missed during the validation process.

**5 Conclusion**

In this paper, we have proposed and used a mixed method for English instrument translation into French using different techniques from different approaches. To the best of our knowledge, the Back translation approach with an adjudicator and monolingual and bilingual tests (B-TAMBiT) is the first of its kind in privacy researches related to SNSs. We are currently using the obtained instrument for data collection. Further inquiry could be made by using the B-TAMBiT with other languages especially locals like Kswahili or Madinka (..) for a better accessibility.

Acknowledgements

We would like to express our sincere thanks to the translator and respondents of the tests for their time and availability. A special mention to our adjudicator for his valuable time.

References

Campbell, A. J. (1997). Relationship marketing in consumer markets: A comparison of managerial and consumer attitudes about information privacy. *Journal of Direct Marketing*, *11*(3), 44-57.

Kante, M., Chepken, C., & Oboko, R. (2017). Methods for translating ICTs' survey questionnaire into French and Bambara. *Knowledge and Innovation for Social and Economic Development*.

Kokolakis, S. (2017). Privacy attitudes and privacy behaviour: A review of current research on the privacy paradox phenomenon. *Computers & security*, *64*, 122-134.

Krasnova, H., Spiekermann, S., Koroleva, K., & Hildebrand, T. (2010). Online social networks: Why we disclose. *Journal of information technology*, *25*(2), 109-125.

Levy, Y., & Ellis, T. J. (2006). A systems approach to conduct an effective literature review in support of information systems research. Informing Science, 9, 181–211. doi: 10.1049/cp.2009.0961

Maneesriwongul, W., & Dixon, J. K. (2004). Instrument translation process: a methods review. Journal of advanced nursing, 48(2), 175–186.

McKnight, D. H., Choudhury, V., & Kacmar, C. (2002). Developing and validating trust measures for e-commerce: An integrative typology. Information systems research, 13(3), 334–359.

Moon, Y. (2000). Intimate exchanges: Using computers to elicit self-disclosure from consumers. Journal of consumer research, 26(4), 323–339.

Pan, Y., & de La Puente, M. (2005). Census bureau guideline for the translation of data collection instruments and supporting materials: Documentation on how the guideline was developed. Survey Methodology, 6.

Sinaiko, H. W., & Brislin, R. W. (1973). Evaluating language translations: Experiments on three assessment methods. Journal of Applied Psychology, 57(3), 328.

Skinner, E. A. (1996). A guide to constructs of control. Journal of personality and social psychology, 71(3), 549.

Webster, J., & Watson, R. T. (2002). Analyzing the past to prepare for the future: Writing a literature review. MIS quarterly, xiii–xxiii.

Wirth, J. (2018). Dependent variables in the privacy-related field: A descriptive literature review.